\documentclass[twocolumn,groupeaddress,showpacs,aps,pra]{revtex4}

\newcommand{\dd}{\mathrm d}

\newcommand{\ii}{\mathrm i}

\newcommand{\calO}{\mathcal O}

\newcommand{\eff}{G}

\newcommand{\loc}{{\rm loc}}

\usepackage{epsf}
\usepackage{amsmath}
\usepackage{amsfonts}
\usepackage{amssymb}
\usepackage{bm}
\usepackage{bbm}
\usepackage{nicefrac}
\usepackage{color}
\usepackage{pifont}
\usepackage{graphicx}
\usepackage{enumerate}
\usepackage{dcolumn}
\usepackage{maybemath}

\begin{document}

\title{Gravitational Correction to Vacuum Polarization}

\author{U.~D.~Jentschura}

\affiliation{Department of Physics, Missouri University of Science and
Technology, Rolla, Missouri 65409, USA}

\affiliation{National Institute of Standards and Technology,
Physics Division, Gaithersburg, Maryland 20899, USA}

\begin{abstract}
We consider the gravitational correction to (electronic) vacuum polarization in
the presence of a gravitational background field. The Dirac propagators for the
virtual fermions are modified to include the leading gravitational correction
(potential term) which corresponds to a coordinate-dependent fermion mass. The
mass term is  assumed to be uniform over a length scale commensurate with the
virtual electron-positron pair. The on-mass shell renormalization condition
ensures that the gravitational correction vanishes on the mass shell of the
photon, i.e., the speed of light is unaffected by the quantum field theoretical
loop correction, in full agreement with the equivalence principle.  Nontrivial
corrections are obtained for off-shell, virtual photons.  We compare our
findings to other works on generalized Lorentz
transformations and combined quantum-electrodynamic gravitational
corrections to the speed of light which have recently appeared in
the literature.
\end{abstract}


\maketitle

%
%
\section{Introduction}
\label{sec1}

The speed of light, in curved space-time,
is not as ``constant'' as one would otherwise imagine.
The curvature of space-time, according to classical general
relativity (see Appendix~\ref{appa}), 
acts as a refractive medium 
(without dispersion), giving rise to an effective 
change in the speed of light 
(as seen from a global, not local, coordinate system),
which reads as 
\begin{equation}
\label{classres}
\frac{\Delta c}{c_0} 
= 2 \, \frac{\Phi_G(\vec r)}{c_0^2} 
= (1+\gamma) \, \frac{\Phi_G(\vec r)}{c_0^2} < 0 \,.
\end{equation}
Here, $\Phi_G(\vec r)$ is the gravitational potential,
normalized to zero for two very distant objects,
and the $\gamma$ parameter is introduced 
(for Einsteinian gravity, we have $\gamma=1$,
see Refs.~\cite{Wi2006,Wi2014}).
Throughout this article, we set $c_0 = 299\,792\,458\, {\rm m}/{\rm s}$ 
equal to the speed of light as consistent with the 
Einstein equivalence principle, which states that 
space-time is locally flat.
The speed-of-light parameter $c_0$
is canonically set equal to unity in 
an appropriate unit system.
The time delay formula~\eqref{classres} is valid to first 
order in the gravitational coupling constant (Newton's constant) $G$.
The concomitant slow-down of light is known 
as the Shapiro time delay~\cite{Sh1964,ShEtAl1968,Sh1999}.
One of the most precise tests has been accomplished with 
the Cassini spacecraft in superior conjunction on its way 
to Saturn~\cite{BeIeTo2003};
it involves Doppler tracking using both X-band ($7175\,{\rm MHz}$)
as well as Ka-band ($34316\,{\rm MHz}$) radar.

At high energy, the dispersion relation for a 
massive particle is not different from that for photons,
$E = \sqrt{\vec p^2 \, c_0^2 + m^2 \, c_0^4}
\approx |\vec p|\, c_0 = \hbar \, |\vec k| \, c_0$,
where $E$ is the energy,
$\vec p$ is the momentum and $\vec k$ is the 
wave vector of the (light or matter) wave.
The modification~\eqref{classres} affects the speed of propagation
for photons as well as highly energetic neutrinos.
For the central field of the Sun, we
have $\Phi_G(\vec r) = -G \, M_\odot/r$ where $M_\odot$ is the
Sun's mass. In general, $\Phi_G$ is negative,
implying that light is slowed down due to the bending
of its trajectory caused by space-time curvature.

Recently, in Ref.~\cite{Fr2014}, it has been claimed that
an additional quantum electrodynamic (QED) correction 
to the result~\eqref{classres} exists, which is of the 
functional form
\begin{equation}
\label{quantcorr}
\frac{\delta c_\gamma}{c_0} = \chi \, \alpha \,
\frac{\Phi_G(\vec r)}{c_0^2} < 0\,,
\end{equation}
where $\alpha$ is the fine-structure constant,
and $\chi$ is a constant coefficient.
For details of the arguments which led Franson
to his result given in Eq.~\eqref{quantcorr}, 
we refer the reader to Sec.~3 of Ref.~\cite{Fr2014}.
Essentially, Franson~\cite{Fr2014} evaluates the vacuum-polarization 
correction for photons (on shell) in the gravitational 
field, using a partially noncovariant formalism
[the photon energy $E$ is used as a noncovariant 
variable in the propagators; see Eq.~(13) ff.~of Ref.~\cite{Fr2014}
for details of Franson's considerations].
It is known from quantum electrodynamic
bound-state calculations that even a slight noncovariance 
in the regularization scheme can induce spurious terms~\cite{PeSaSu1998};
some scrutiny should thus be applied.
Using his calculational scheme,
Franson comes to the conclusion that the speed of 
{\em photons}
[hence the subscript $\gamma$ in Eq.~\eqref{quantcorr}]
is altered due to the gravitational correction 
to the electron-positron propagators that enter 
the vacuum-polarization loop calculation.
For the coefficient $\chi$, 
the following result has been indicated in Ref.~\cite{Fr2014},
\begin{equation}
\label{coeffres}
\chi = \frac{9}{64}
\qquad
\mbox{(according to Ref.~\cite{Fr2014}).}
\end{equation}
The decisive point of the analysis 
presented in Ref.~\cite{Fr2014} is that the 
effect described by Eq.~\eqref{quantcorr} is claimed to 
affect only photons, not neutrinos,
thus slowing the photons in comparison to the 
neutrinos (and other massive fermions).
According to Ref.~\cite{Fr2014},
the propagation of otherwise massless 
photons is influenced by the electron-positron
(light fermion) vacuum-polarization effect at one-loop
order, that of fermions is not.

A different {\em ansatz} for a modification of local Lorentz
transformations stems
from the work of Vachaspati~\cite{Va2004},
who claims that in addition to ``electromagnetic time'',
one can define an ``absolute time'' (on the level of 
special relativity), which transforms according to a
modified Lorentz transformation (referred to here as  
the Vachaspati transformation), and 
which, according to Ref.~\cite{Va2004}, is  claimed to 
be compatible with muon lifetime and Michelson--Morley experiments
(see Appendix~\ref{appb}).
The Vachaspati transformation also leads to 
a ``speed of light'' parameter which is dependent on the
inertial frame.

Here, we aim to investigate three sets of questions:
{\em (i)} Is the result given in Eq.~\eqref{quantcorr} 
compatible with all other astrophysical observations
recorded so far in the literature? 
What bounds can be set for the $\chi$ parameter given in 
Eq.~\eqref{coeffres}?
{\em Irrespective of the value of $\chi$,}
what changes would result from a
hypothetical quantum modification of the 
speed of light, induced according to the functional 
form Eq.~\eqref{quantcorr},
for the description of other 
physical phenomena? In particular, how would we describe neutrinos 
in strong gravitational fields, where according to Ref.~\cite{Fr2014},
they propagate faster than the speed of light, even at high 
energy? How would the result
given in Eq.~\eqref{quantcorr} affect the
Schiff conjecture~\cite{Wi2001,Wi2006,Wi2014}?
{\em (ii)} The next question then is
whether  the modification given in  Eq.~\eqref{quantcorr}
{\em exists at all.} In Sec.~\ref{sec3}, we investigate
whether or not the calculations reported in 
Ref.~\cite{Fr2014}, which lead to the quantum effect~\eqref{quantcorr},
stand the test of a fully covariant formulation of 
the gravitational corrections to vacuum polarization,
where the virtual fermions in the loop 
are subject to gravitational interactions.
Our calculation is restricted to an analysis 
of the electron-positron loop insertion into the 
photon propagator, which is the subject of Ref.~\cite{Fr2014},
and does not treat all possible quantum corrections to 
the photon propagator into account.
The analysis in Sec.~\ref{sec2} 
thus covers a much more general scope and answers general questions
regarding a modification of the speed of light in gravitational
fields, induced according to Eq.~\eqref{quantcorr}, while the 
analysis in Sec.~\ref{sec3} only covers the vacuum-polarization 
loop with fermion propagators subject to gravitational 
interactions.
{\em (iii)} In the context of atomic physics,
what phenomenological consequences
will result from the gravitational correction to the 
off-shell (virtual) photon propagator?
This is briefly discussed in Sec.~\ref{sec4}.
The first set of questions also has relevance for the 
work of Vachaspati~\cite{Va2004}.
Conclusions are reserved for Sec.~\ref{sec5}.

%
%
\section{Quantum Effects and Speed of Light}
\label{sec2}

%
%
\subsection{Quantum correction and Shapiro time delay}
\label{sec2A}

Because the quantum correction~\eqref{quantcorr} is 
conjectured to be induced by a virtual 
loop consisting of electrons and positrons
lifted from the quantum vacuum,
its existence is not excluded by classical
theory, i.e., beyond the validity of the 
original (purely classical) general theory of relativity formulated 
by Einstein and Hilbert~\cite{Ei1915,Hi1915,Ei1916}.
The delay induced by the conjectured 
modification Eq.~\eqref{quantcorr} for light rays 
propagating from the Large Magellanic Cloud 
is claimed to be in agreement~\cite{Fr2014} with the observed early 
arrival time of the (still somewhat mysterious) 
early neutrino burst under the Mont Blanc recorded 
in temporal coincidence with 
the SN1987A supernova~\cite{DaEtAl1987}.
Essentially, the paper~\cite{Fr2014} claims that the 
apparent superluminality of the ``early'' neutrino 
burst could be due to a quantum 
electrodynamic effect which slows down light in 
comparison to the neutrinos, 
in strong gravitational potentials,
with a delay induced according to Eq.~\eqref{quantcorr}.

\begin{figure}[t!]
\begin{center}
\begin{minipage}{1.0\linewidth}
\includegraphics[width=0.99\linewidth]{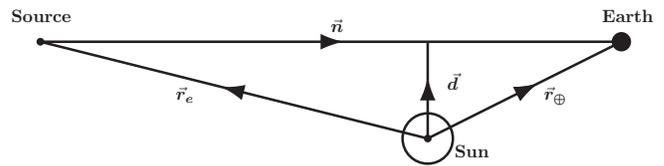}
\caption{\label{fig1} Geometry for Eq.~\eqref{std}.}
\end{minipage}
\end{center}
\end{figure}

However, this result should be compared to 
other precision measurements of time delays induced 
by space-time curvature, such as the 
Shapiro time delay~\cite{Sh1964,ShEtAl1968,Sh1999}.
The time delay due to the refractive index of curved space 
leads to the following formula for 
a light ray or radar wave
as it bounces back from an object close to superior conjunction
[see Eq.~(49) of Ref.~\cite{Wi2006} and Fig.~\ref{fig1}],
\begin{equation}
\label{std}
\delta t = 2 \, \left(1 + \gamma \right) \,
\frac{G \, M_\odot}{c_0^3} \,
\ln\left( \frac{(r_\oplus + \vec r_\oplus \cdot \vec n) \,
(r_e - \vec r_e \cdot \vec n)}{d^2} \right) \,.
\end{equation}
Here, $\vec r_e$ is the vector from the Sun to the 
source (e.g., the Cassini spacecraft), 
$\vec r_\oplus$ is the unit vector from the Sun to the Earth, 
while $\vec n$ is the vector from the source to the Earth,
and $d$ is the distant of closest approach of the light ray 
as it travels from the Earth to the source and back.
The parameter $\gamma$ is used in order to 
describe potential deviations from the classical prediction.

The formula~\eqref{std} is obtained on the basis of 
the classical result~\eqref{classres}.
For details of the derivation,
we refer to Chap.~4.4 on page 196~ff.~of Ref.~\cite{OhRu1994}
and exercise 4.8 of page 161 of Ref.~\cite{Pa2010}.
The quantum ``correction'' given in~\eqref{quantcorr}
has the same functional form as the classical result~\eqref{classres}
but adds a correction to the prefactor.
If we assume the explicit numerical result 
given in Eq.~\eqref{coeffres} to be valid, then this 
leads to a $\gamma$ coefficient different from unity,
\begin{equation}
\gamma - 1 = \chi \, \alpha = 
\frac{9 \, \alpha}{64} = 1.03 \times 10^{-3} \,.
\end{equation}
However, the 
{\color{black} result of the Cassini observations~\cite{BeIeTo2003}} reads as follows,
\begin{equation}
\label{experimental}
\gamma - 1 = ( 2.1 \pm 2.3 ) \times 10^{-5} \,.
\end{equation}
The claim~\eqref{coeffres} thus is in a $44.8 \sigma$ disagreement 
with the experimental result~\eqref{experimental},
which is otherwise consistent with zero. Unless
the authors of Ref.~\cite{BeIeTo2003} have overlooked a significant 
source of systematic error, the effect described by
Eq.~\eqref{coeffres} thus is in severe disagreement 
with experiment.
Finally, we should remark that
the $\gamma$ parameter also enters the expression for
the light deflection formula around a central gravitational centre.
{\color{black} An analysis~\cite{ShDaLeGr2004}}  of almost 2
million very-long baseline (VLBI) observations of 541 radio 
sources, made by 87 VLBI sites yields the bound
\begin{equation}
\delta\gamma = (-1.7 \pm 4.5) \times 10^{-4} \,,
\end{equation}
which also is in disagreement with the claim~\eqref{quantcorr}.
According to Refs.~\cite{LaPL2009,LaPL2011},
all current VLBI data together yield a value of
$\delta\gamma = (−0.8 \pm 1.2) \times 10^{−4}$,
compatible with zero.

Alternatively, we can convert the result~\eqref{experimental}
into a bound for the $\chi$ coefficient,
\begin{equation}
\label{bound}
\chi = ( 2.9 \pm 3.2 )  \times 10^{-3} \,,
\end{equation}
consistent with zero. However, quantum effects of the 
functional form~\eqref{quantcorr}, 
but with a numerically small coefficient compatible with 
the bound~\eqref{bound}, cannot be excluded at present.

%
%
\subsection{Fermion wave equation}
\label{sec2B}

Let us analyze the problem of a fermion wave equation
for a local Lorentz frame in which 
photons propagate slower than high-energy fermions.
We remember that the Lorentz violation induced by Eq.~\eqref{quantcorr} actually is 
quite subtle; the effect is not excluded by classical
physics and vanishes globally in the absence 
of gravitational interactions, i.e., 
it does not perturb the speed of light in globally 
flat (Minkowski) space-time.
In order to write a wave equation describing fermions,
we have to carefully distinguish between 
the flat-space speed of light {\color{black} $c_0$} (in the absence of gravitational
interactions), 
the classical ``correction'' $\Delta c$
(which is compatible with the Einstein equivalence principle
and does not preclude the existence of the local Minkowskian 
frame of reference),
and the quantum correction $\delta c_\gamma$ given in Eq.~\eqref{quantcorr},
which changes the speed of light in a ``local'' reference 
frame to 
\begin{equation}
c_\loc = {\color{black} c_0} + \delta c_\gamma = 
{\color{black} c_0} - |\delta c_\gamma| \,.
\end{equation}
We recall that, physically, the speed of light is the speed
which describes the propagation of the transverse components of the
electromagnetic field, which enter the Maxwell equations.
(The necessity of a
careful separation of transverse and longitudinal components has recently been
highlighted in a consideration of the photon wave functions, given in
Ref.~\cite{Mo2010}.)

As already stated in Sec.~\ref{sec1},
according to Ref.~\cite{Fr2014},
the modification~\eqref{quantcorr} is supposed to 
slow down photons in strong gravitational 
fields, not neutrinos or electrons.
Let us therefore investigate the question of a 
correct equation to describe fundamental fermions 
in strong gravitational fields (deep potentials),
on the basis of a (possibly generalized) Dirac equation.
One possibility is to postulate that
the local Lorentz transformation has to be modified to include the local
quantum modification of the speed of light, while the formalism of classical
general relativity is unaltered by the quantum modification. 
Let us also assume that the ``local Lorentz transformation'',
under the presence of the quantum correction~\eqref{quantcorr},
is formulated to be the transformation which preserves the 
light element
\begin{equation}
\dd x^\mu \, \dd x_\mu =
c_\loc^2 \dd t^2 - \dd \vec r^{\,2} = 0\,,
\end{equation}
where $\dd x^\mu = ( c_\loc \, \dd t, \dd \vec r)$ is a
space-time interval,
and $c_\loc$ is the speed of light in the local coordinate system.
The correction $\Delta c$ given in Eq.~\eqref{classres}
is compatible with the Einstein equivalence principle
of a locally flat space-time and therefore does not 
change the Dirac equation (with parameter $c$)
in the usual Dirac equation
for fermion wave packets, but the quantum correction $\delta c_\gamma$, 
given in Eq.~\eqref{quantcorr}, leads to the 
replacement ${\color{black} c_0} \to c_\loc$.

According to Eq.~\eqref{quantcorr},
high-energy fermions are faster than light rays at high energy, by an 
offset $|\delta c_\gamma|$, making them effectively superluminal,
thus leading to an explanation for the early neutrino 
burst from the supernova 1978A (see Refs.~\cite{DaEtAl1987,Fr2014}). 
The preferred way to describe highly energetic fermions (neutrinos) 
which travel faster than {\color{black} light} is via the tachyonic Dirac 
equation~\cite{ChHaKo1985}, which in the local reference frame 
reads as 
\begin{equation}
\label{tach}
\left( \ii \hbar \, \gamma^\mu \, \frac{\partial}{\partial x^\mu} - 
\gamma^5 m c_\loc \right) \, 
{\color{black} \psi}(t, \vec r) = 0 \,,
\end{equation}
{\color{black} where ${\color{black} \psi}(t, \vec r)$ is the fermion wave function.}
The projector sums for the tachyonic 
spinor solutions have recently been investigated
in Refs.~\cite{JeWu2012epjc,JeWu2014}. 
The main problem 
here does not lie in the tachyonic equation,
but in the description of highly energetic 
neutrinos because of their uniform velocity 
offset $|\delta c_\gamma|$ at high energy from photons.
This offset prevents them from 
reaching the photon mass shell in the local 
coordinate system. To see this, let us note the particles
described by Eq.~\eqref{tach} fulfill the 
dispersion relation
\begin{subequations}
\begin{align}
\label{tachdisp}
E =& \; \sqrt{ \vec p^{\,2} \, c^2_\loc - (m c^2_\loc)^2} \,,
\\[0.77ex]
E =& \; \frac{m c_\loc^2}{ \sqrt{ v^2/c_\loc^2 - 1 }} \,,
\\[0.77ex]
|\vec p| =& \; \frac{m v}{ \sqrt{ v^2/c_\loc^2 - 1 }} \,,
\end{align}
\end{subequations}
where {\color{black} $v \approx c_0 > c_\loc$} is the propagation speed
of highly energetic neutrinos, required for the explanation 
of the early arrival time of the neutrinos according 
to Ref.~\cite{Fr2014}.
The energy can thus be expressed as 
\begin{subequations}
\begin{align}
\label{EE}
E =& \; \frac{m c_\loc^2}{\sqrt{ v^2/c_\loc^2 - 1 }} \approx
\frac{m \, c_\loc^{5/2}}{\sqrt{ 2 |\delta c_\gamma| }} \,,
\\[0.77ex]
v =& \; c_\loc  + |\delta c_\gamma| \approx {\color{black} c_0}  \,.
\end{align}
\end{subequations}
We are now in a dilemma: On the one hand,
the energy of a highly energetic neutrino
is not bounded from above, but even for a neutrino
traveling {\em exactly} at the speed of light 
{\color{black} $v = c_0$},
the right-hand side of Eq.~\eqref{EE} only contains the 
fixed parameters $ c_\loc$ and $|\delta c_\gamma|$.
Hence, the only way to make Eq.~\eqref{EE} compatible with 
Eq.~\eqref{tach} is to assume a universal mass ``running''
of the tachyonic mass parameter in Eq.~\eqref{tach}, 
linear with the energy scale, of the functional form
\begin{equation}
m \to m(E) \propto E = \frac{\sqrt{2 |\delta c_\gamma|}}{c_\loc^{5/2}} \, E \,.
\end{equation}
It thus becomes clear that the mere existence of 
a ``local'' gravitational quantum correction 
of the functional form~\eqref{quantcorr} 
would induce severe problems in the description of
high-energy fermions in local reference frames in 
strong gravitational fields (``deep potentials'').
In other scenarios of Lorentz breaking mechanisms
in local reference frames~\cite{BaKo2006,Ba2009,KoTa2011},
the Lorentz-breaking terms are not required to 
run with the energy scale.
The same is true for small Lorentz-violating 
admixture terms to Dirac equations in free 
space~\cite{CoKo1997,DiKoMe2009,KoMe2012}.

As a final remark, let us note that 
according to Ref.~\cite{Fr2014}, high-energy neutrinos 
would be traveling faster than {\rm light}, but 
not faster than electrons. 
Hence, the analogue of Cerenkov radiation emitted
by neutrinos, namely, the reaction
$\nu \to \nu + e^+ + e^-$ cannot occur;
according to Ref.~\cite{CoGl2011}, 
this process constitutes the main decay channel of 
{\color{black} tachyonic neutrinos.}
Genuine Cerenkov radiation 
$\nu \to \nu + \gamma$ is suppressed for the 
electrically neutral neutrinos and must proceed 
via a $W$ loop. The slow-down of light in comparison to 
high-energy fermions according to Eq.~\eqref{quantcorr}, 
though, would lead to Cerenkov radiation from 
highly energetic charged leptons [e.g.,
synchrotron losses at the Large 
{\color{black} Electron--Positron Collider} (LEP)].
Within the models studied in Refs.~\cite{KlSc2008,Al2009,HoLePhWe2009},
rather stringent bounds have been obtained for 
certain Lorentz-violating parameters.
All of these results, though, are model dependent. 
E.g., the dispersion relation $E = p \, v_\nu$,
assumed in Ref.~\cite{CoGl2011}, is different from the 
dispersion relation that is generally assumed for 
tachyonic neutrinos [see Eq.~\eqref{tachdisp} here in the 
paper and independently Ref.~\cite{ChHaKo1985}].
The Lorentz violation induced 
by the slow-down of light due to a radiative correction
proposed in Ref.~\cite{Fr2014} is quite subtle;
however, the functional form~\eqref{quantcorr} 
allows for a direct model-independent comparison with bounds on the 
$\gamma$ parameter introduced in Eq.~\eqref{classres},
as discussed in Sec.~\ref{sec2A}.

%
%
\subsection{Equivalence principle and Schiff conjecture}
\label{sec2C}

The Schiff conjecture (see Sec.~2.2.1 of Ref.~\cite{Wi2001})
is connected with two different
forms of the equivalence principle, namely, the weak equivalence principle and
the Einstein equivalence principle. Originally,
Newton stated that the property of a body called
``mass'' (``inertial mass'') is proportional to the ``weight'' (which enters
the gravitational force law), a principle otherwise known 
as the ``weak equivalence principle'' (WEP).
The Einstein equivalence principle (EEP) states that {\em (i)} WEP is valid,
{\em (ii)} the outcome of any local non-gravitational experiment is independent
of the velocity of the freely-falling reference frame in which it is performed
(local Lorentz invariance, LLI),
{\em (iii)} the outcome of any local non-gravitational experiment is
independent of where and when in the universe it is performed
(local position invariance, LPI).

It is obvious to realize that the existence of a local modification of the
speed of light in deep gravitational potentials according to~\eqref{quantcorr}
would lead to a (very slight, but noticeable) violation of point {\em (iii)} of
the EEP. Namely, because the shift $\delta c_\gamma$ affects
only photons, not neutrinos or electrons (according to Ref.~\cite{Fr2014}), 
one could measure the local propagation velocity of high-energy 
fermion versus photon wave packets. The former propagate at 
velocity $c_0$ in a local reference frame, whereas 
the latter are affected by the correction~$\delta c_\gamma \propto \Phi_G$.
The potential $\Phi_G$ depends on the position in the Universe
where the experiment is performed 
(for reference values of $\Phi_G$ in different regions,
see Table~1 of Ref.~\cite{Fr2014}).

According to Sec.~2.2.1 of Ref.~\cite{Wi2001}, the 
Schiff conjecture states that
for self-consistent theories of gravity, WEP necessarily embodies EEP; the
validity of WEP alone guarantees the validity of local Lorentz and position
invariance, and thereby of EEP.  The question of whether the correction $\delta
c_\gamma$ violates the WEP is a matter of interpretation because $\delta
c_\gamma$ affects only massless objects, namely, photons; it is, as already
emphasized, a {\em quantum} effect which goes beyond the scope of classical
mechanics in which the weak equivalence principle was first formulated
(in its original form by Newton).  

One could perform a thought experiment and enter a region of deep gravitational
potential with three freely falling, propagating wave packets, one describing a
photon, the others describing a very highly energetic neutrino and a very
highly energetic electron, respectively. The latter two propagate at a velocity
(infinitesimally close to) $c_0$. If a correction of the form $\delta c_\gamma$
exists, then photons will have been decelerated to a velocity $c_0 - |\delta
c_\gamma|$ within the deep gravitational potential, whereas {\em both} fermions
will have retained a velocity (infinitesimally close to) $c_0$.  If we regard
the photons as particles (the photon being a concept introduced into physics
after the WEP was first introduced by Newton), then we could argue that a
``force'' must have acted onto the photon, causing deceleration, even though
the particles were in free fall. This might indicate a violation of the WEP but
only if the photon were regarded as a normal ``particle'' in the sense of
Newton's idea (which is not fully applicable because of the vanishing rest mass
of the photon).  Alternatively, we could interpret any change in velocities
relative to the local speed of light as an ``acceleration'' and thus interpret
the faster propagation of the electrons and neutrinos in comparison to the
photon within the region of deep gravitational potential as the result of a
force which must have acted on the fermions.  {\em Both} neutrinos and
electrons retain a velocity very close to $c_0$ and have thus been accelerated
by the same velocity $|\delta c_\gamma|$; because of their different rest mass,
the force acting on them must have been different, thus violating the WEP.

Today, one canonically understands the WEP as not being tied to ``massive''
objects, stating that free-fall at a given point in space-time is the same for
all physical systems, and that photons, electrons, and neutrinos in a
gravitational potential all act as if they are in the same accelerated
coordinate frame.  In that sense, if a theory predicts that gravitational
potentials make the local photon velocity different from the local limiting
velocity of high-energy massive particles, then that theory violates the WEP. 

Thus, depending on the interpretation, one might conclude that Schiff's
conjecture holds true, in the sense that the correction~\eqref{quantcorr}
violates {\em both} the WEP as well as the EEP.  The caveat must be stated
because strictly speaking, photons do not have a rest mass, and thus, the WEP
in the original formulation is not fully applicable.  One should also bear in
mind that slight violations of fundamental laws and symmetries of nature are
being discussed and all we can do is establish bounds for violating
parameters~\cite{BaKo2006,Ba2009,KoTa2011,CoKo1997,DiKoMe2009,KoMe2012}.  For
the scenario studied by Vachaspati (see Appendix~\ref{appb} and
Ref.~\cite{Va2004}), the violations of the EEP and the WEP would be of order
unity; the ``light speed measured in absolute time'' can be different from the
``light speed measured in electromagnetic time'', depending on the relative
velocity of the moving frames $v_A$.

%
%
\section{Dirac Equation and Gravitational Coupling}
\label{sec3}

The far-reaching consequences of any correction of the 
form~\eqref{quantcorr} to the speed of light in deep gravitational
potentials together with the bound formulated in Eq.~\eqref{bound}
for the $\chi$ coefficient stimulate a recalculation of the 
leading gravitational correction to vacuum polarization,
supplementing the analysis of Ref.~\cite{Fr2014}.
Recently, the gravitationally coupled Dirac equation has been 
investigated~\cite{Je2013,JeNo2013pra,Je2014dirac,Je2014pra},
with particular emphasis on the Dirac--Schwarzschild problem,
which is the equivalent of the Dirac--Coulomb problem 
for electrostatic interactions and 
describes a particle bound to a central gravitational field.
From now on, for the remainder of this article,
we revert to natural units with $\hbar = c_0 = \epsilon_0 = 1$,
because we no longer consider a conceivable ``correction''
of the form~\eqref{quantcorr}.
In leading order, the Hamiltonian which governs the 
gravitational interaction is given by [see Eq.~(12) of 
Ref.~\cite{Je2013}]
\begin{align}
\label{Hgrav}
H =& \; \vec\alpha \cdot \vec p + \beta \, m \,  w(r) \,,
\\[0.133ex]
w \approx & \; 1 - \frac{r_s}{2 r} 
= 1 - \frac{G \, M}{r}
= 1 + \Phi_G \,,
\end{align}
where $r_s = 2 G M$ is the Schwarzschild radius.  Here, $r$ is the
Eddington coordinate in the Eddington form~\cite{Ed1924} of the Schwarzschild
metric, which however is equal to the radial coordinate in the original
Schwarzschild metric in the limit $r \to \infty$ (i.e., in the limit of a weak
gravitational field). We use the vector of Dirac $\vec\alpha$ matrices, and the
$\beta$ matrix, in the standard representation~\cite{ItZu1980}.

After a Foldy--Wouthuysen transformation,
the Hamiltonian~\eqref{Hgrav} takes the form (in the leading 
order in the relativistic expansion)
\begin{equation}
\label{schr}
H \approx \beta 
\left(m + \frac{ \vec p^{\,2} }{2 m} - \frac{r_s}{2 r} \right) =
\beta \left(m + \frac{ \vec p^{\,2} }{2 m} + \Phi_G \right) \,.
\end{equation}
Here, the $\beta$ matrix describes the particle-antiparticle 
symmetry~\cite{JeNo2013pra}, while the latter form shows that 
the gravitational potential can be 
inserted into the Schr\"{o}dinger equation 
``by hand'' in the leading order (the somewhat nontrivial
relativistic corrections involve the gravitational {\em Zitterbewegung}
term, and the gravitational spin-orbit coupling~\cite{JeNo2013pra}).

The leading gravitational term in Eq.~\eqref{Hgrav}, 
in the fully relativistic formalism, corresponds to 
a position-dependent modification of the Dirac mass 
of the electron, which is present only if one departs 
from the local Lorentz frame (locally flat space-time) 
and aims to describe the Dirac particle globally, in the curved 
space-time. 
Defining the effective mass $m_G$ of the electron as
\begin{equation}
m_{\eff} = m \, w(r) \approx
m \left( 1 + \Phi_G \right) \,,
\end{equation}
one can carry out the calculation of the 
vacuum polarization insertion as described 
in the literature.
One possibility is to use the covariant 
formalism described in Chap.~7 of Ref.~\cite{ItZu1980}, which 
relies on a Feynman parameter integral.
A recent, particularly clear formulation given in 
Sec.~5 of Ref.~\cite{InMoSa2014} clarifies that the 
additional mass terms introduced in Pauli--Villars 
regularization do not affect the 
calculation of the vacuum-polarization tensor, which 
depends only on the physical, local, effective mass of the 
electron. An alternative possibility is given 
in Chap.~113 of Ref.~\cite{BeLiPi1982vol4}, where
a subtracted dispersion relation is used in order to 
circumvent parts of the problems associated with 
regularization and renormalization,
and leads to a dispersion integral which starts
at the pair production threshold $(2 m_{\eff})^2$.
The result of all these approaches invariantly reads as 
follows, in terms of a modification of the 
photon propagator $D_{\mu\nu} = g_{\mu\nu}/k^2$,
\begin{equation}
\label{feyn}
\frac{g_{\mu\nu}}{k^2} \to
\frac{g_{\mu\nu}}{k^2 \, [ 1 + {\overline \omega}^R(k^2) ]}
\,,
\qquad 
k^2 = \omega^2 - \vec k^2 \,.
\end{equation}
A straightforward application of the formalism of 
covariant quantum electrodynamics then leads to 
the renormalized (superscript $R$) vacuum-polarization
insertion, written in terms of the effective mass $m_G$ of the electron,
\begin{equation}
\label{R}
{\overline \omega}^R(k^2) = 
\frac{\alpha k^2}{3 \pi} \,
\int\limits_{4 m_{\eff}^2}^\infty 
\frac{\dd k'^2}{k'^2} \, 
\frac{1 +  2 m_{\eff}^2/k'^2}{k'^2 - k^2} 
\sqrt{ 1 - \frac{4 m_{\eff}^2}{k'^2} } \,.
\end{equation}
We note that ${\overline \omega}^R(k^2)$ 
vanishes for $k^2 = \omega^2 - \vec k^2 = 0$,
thus leaving the speed of light of on-shell photons invariant.
For $k^2 \neq 0$ (off-shell, virtual photons),
we note the asymptotic behavior
\begin{subequations}
\label{off}
\begin{align}
{\overline \omega}^R(k^2) =& \;
\dfrac{\alpha}{15 \pi} \dfrac{k^2}{m_{\eff}^2} + \calO(k^4) \,, 
\qquad
k^2 \to 0 \,, 
\\[2ex]
{\overline \omega}^R(k^2) =& \;
-\dfrac{\alpha}{3 \pi} \, 
\ln\left( - \dfrac{k^2}{m_{\eff}^2} \right) + \dfrac{5\alpha}{3\pi} 
+ \calO\left(\frac{\ln(-k^2)}{k^2} \right) \,, 
\nonumber\\[0.133ex]
& \; k^2 \to \infty \,.
\end{align}
\end{subequations}
These are in principle familiar formulas (see Chap.~7 of Ref.~\cite{ItZu1980}),
and we identify the leading gravitational effect on 
vacuum polarization to be given by the 
gravitationally corrected mass.  The conclusions of
Ref.~\cite{Fr2014}, and the result~\eqref{quantcorr}, can thus be traced to an
inconsistent evaluation of the vacuum polarization integral, which relies on a
relativistically noncovariant formulation [see the
discussion surrounding Eq.~(6) of Ref.~\cite{Fr2014}],
and bears an analogy with similar problems encountered 
in bound-state quantum electrodynamics~\cite{PeSaSu1998}.

%
%
\section{Bound--State Energies}
\label{sec4}

A final word on bound-state energies is in order.  With the mass of the
electron assuming the value $m \to m_{\eff}$, the vacuum polarization potential
(Uehling, one loop), derived from the virtual exchange of space-like Coulomb
photons ($k^2 = -\vec k^2$), is easily derived as
(in units with $\hbar = c_0 = \epsilon_0 = 1$)
\begin{equation}
V_{\rm vp}(\vec r) =
-\dfrac{4 \alpha}{15} \frac{Z \, \alpha}{m_G^2} \, \delta^{(3)}(\vec r) \,,
\end{equation}
where $Z$ is the nuclear charge number.
However, the gravitationally corrected mass also enters the 
Dirac--Coulomb Hamiltonian $H = \vec\alpha \cdot \vec p + 
\beta \, m_G - Z\alpha/r$, where the Dirac matrices are used in the 
standard representation, and $r$ denotes the electron-proton 
distance~\cite{ItZu1980}. By consequence, after a Foldy--Wouthuysen
transformation, the gravitationally corrected mass parameter
$m_G$ also enters the Schr\"{o}dinger wave function,
and the probability density of $S$ states with 
orbital angular momentum $\ell = 0$ at the origin becomes
proportional to $(Z\alpha m_G)^3$. 
The gravitationally corrected energy shift reads as
\begin{equation}
\left< V_{\rm vp}(\vec r) \right> =
-\dfrac{4 \alpha}{15 \pi} \frac{(Z \, \alpha)^4 \, m_{\eff}}{n^3} \,
\delta_{\ell \,0} \,.
\end{equation}
The energy shift is proportional to the effective mass of the electron,
which also enters the Schr\"{o}dinger spectrum $E_n = -(Z\alpha)^2 m_G/(2 n^2)$,
where $n$ is the principal quantum number.
[We recall that the Dirac-$\delta$ potential $\delta^{(3)}(\vec
r)$ is formulated with respect to the central electrostatic potential generated
by the nucleus of charge number $Z$, not the gravitational centre, while $n$
and $\ell$ denote the principal and orbital angular momentum quantum numbers of
the state.] The scaling with the effective mass of the electron thus affects
the vacuum polarization energy shift as much as the leading Schr\"{o}dinger
term and thus does not shift atomic transitions with respect to each other.

The gravitational correction to bound-state energy levels
due to fluctuations of the electron position in the 
gravitational field of the Earth
can easily be estimated as follows.
Namely, the atomic electron coordinate fluctuates over a distance of a Bohr 
radius about the position in the gravitational field. 
If we denote by $\vec R = \vec r_N + \vec r$ the electro coordinate from the 
Earth's centre (with the Earth mass being denoted as 
$M_\oplus$), where $r_N$ is the proton coordinate, then
the fluctuations of the electron about the gravitational
centre of the atom cause an energy shift of the order of 
\begin{equation}
-\frac{G m_e M_\oplus}{ | \vec r_N + \vec r |} + 
\frac{G m_e M_\oplus}{ | \vec r_N | } \sim
\frac{G m_e M_\oplus \, a_0}{ \vec R^2 } = 
2.9 \times 10^{-21} \, {\rm eV} \,.
\end{equation}
This effect influences typical atomic transitions (with transition
frequencies on the order of one eV) at the level of one part in $10^{21}$.

%
%
\section{Conclusions}
\label{sec5}

The main results of the current investigation can be summarized as follows:
Both the Vachaspati transformation (see Appendix~\ref{appa} and
Ref.~\cite{Va2004}) as well as the Franson time delay [see
Eq.~\eqref{quantcorr} and Ref.~\cite{Fr2014}] are in disagreement with the
Einstein equivalence principle (EEP, see the discussion in Sec.~\ref{sec2C}).
The Franson time delay affects only photons, not fermions, is a subtle effect,
and the violation of the EEP due to the 
Franson time delay would be at the quantum level (hence a small correction)
as opposed to the Vachaspati transformation.  Hence, it is warranted to
establish an astrophysical bound on the magnitude of the $\chi$ parameter
introduced in Eq.~\eqref{quantcorr}.  This is done is Sec.~\ref{sec2A}.
Furthermore, the description of fermions in deep gravitational potentials,
under the assumption of a time delay $\delta c_\gamma$ for photons according to
Eq.~\eqref{quantcorr}, is studied in Sec.~\ref{sec2B}.  It is shown that the
description of fermions in such a deep gravitational potential will require a
mass term that ``runs'' with the energy and thus is more problematic than a
superficial look at the ``small'' correction term~\eqref{quantcorr} would
otherwise suggest.

In Sec.~\ref{sec3}, we analyze the leading gravitational correction to vacuum
polarization using a fully covariant formalism and find that, with
on-mass-shell renormalization, the effect can be described by a mass term
modification which depends on the value of the gravitational potential in the
vicinity of the virtual electron-positron pair.  It vanishes on shell and thus
does {\em not} lead to a nonvanishing $\chi$ coefficient in the sense of
Eq.~\eqref{quantcorr}.  Finally, in Sec.~\ref{sec4}, we analyze conceivable
shifts for atomic bound-state levels, caused by off-shell virtual photons in
the vacuum-polarization loops.  We find that the effect,
at least within the approximations employed in 
Sec.~\ref{sec4}, does not shift spectral
lines with respect to each other because it can be absorbed in a prefactor of
the vacuum-polarization term which is also present in the leading
Schr\"{o}dinger binding energy.  Finally, we estimate the leading gravitational
correction to atomic energy levels, which depends on the quantum numbers, in
terms of fluctuations of the electron and nucleus coordinates in the
gravitational field of the Earth, and come to the conclusion that the term
induced by the coordinate fluctuations within 
the binding Coulomb potential is of relative order $10^{-21}$.

%
%
\section*{Acknowledgments}

The author acknowledges helpful conversations with Professor P.~J.~Mohr and
thank A. Migdall for directing our attention to the phenomenological
consequences of the paper by J. D. Franson~\cite{Fr2014}. This research has
been supported by the National Science Foundation (Grants PHY--1068547 and
PHY--1403973).

\appendix

%
%
\section{Global Reference Frame and Speed of Light}
\label{appa}

Let us motivate the Shapiro time delay on the 
basis of the Schwarzschild metric~\cite{Sc1916},
in isotropic form
(Sec.~43 of Chap.~3 of Ref.~\cite{Ed1924}),
\begin{align}
\dd s^2 =& \;
\left( \frac{1 - r_s/(4 r)}{1 + r_s/(4 r)} \right)^2 \, \dd t^2
\\[0.77ex]
& \; -\left( 1 + \frac{r_s}{4 r} \right)^4 \;
\left( \dd r^2 + r^2 \dd \theta^2 +
r^2 \, \sin^2 \theta \, \dd \varphi^2 \right) \,.
\nonumber
\end{align}
We use units with $\hbar = c_0 = \epsilon_0$ for the 
entire Appendix~\ref{appa}. Light travels on a null geodesic, 
with $\dd s^2 = 0$, and so 
\begin{equation}
\left( \frac{1 - r_s/(4 r)}{1 + r_s/(4 r)} \right)^2 \, \dd t^2
-\left( 1 + \frac{r_s}{4 r} \right)^4 \; \dd \vec r^{\,2} = 0 \,.
\end{equation}
One obtains
\begin{align}
\left( \frac{\dd \vec r}{\dd t} \right)^2 
=& \;  \frac{[ 1 - r_s/(4 r) ]^2}{[ 1 + r_s/(4 r) ]^6} =
\left( 1 - 2 \frac{r_s}{r} + \calO\left( \frac{1}{r^2} \right) \right) \,.
\end{align}
We now consider the limit of large distance $r$.
Using the relation $r_s = 2 G M$,
the local speed of light, expressed in terms of the 
global coordinates, is 
\begin{align}
\left| \frac{\dd \vec r}{\dd t} \right|
= 1 - \frac{2 G M}{r} = 1 + 2 \, \Phi_G(\vec r) \,,
\end{align}
where we identity $\Phi_G(\vec r) = - G M/r$ with the gravitational 
potential. One can easily generalize the 
derivation 
[see Chap.~4.4 on page 196 ff.~of Ref.~\cite{OhRu1994},
Eq.~(4.43) of Ref.~\cite{Pa2010},
and Sec.~4.5.2 as well as the discussion on page 160,
and exercise 4.8 on p. 161 of Ref.~\cite{Pa2010},
{\color{black} as well as Ref.~\cite{As2002}}].
The effect is known as the Shapiro time delay~\cite{Sh1964,ShEtAl1968,Sh1999}.
The application to the travel time of particles
stemming from the SN1987A supernova is discussed in
Refs.~\cite{Lo1988,KrTr1988}.

%
%
\section{Vachaspati Transformation}
\label{appb}

Vachaspati~\cite{Va2004} distinguishes between ``absolute time'' $t_A$ 
and ``electromagnetic time'' $t_E$.
The Lorentz--Vachaspati transformation
resembles the Lorentz transformation, 
but with a variable ``speed-of-light parameter'' $u_0$, 
which, in the primed system, transforms into $u'_0$.
\begin{subequations}
\label{LORVA}
\begin{align}
u'_0 \, t'_A =& \; \gamma_A ( u_0 \, t_A + \beta_A \, x ) \,,
\\[2ex]
x' =& \; \gamma_A ( x + \beta_A u_0 \, t_A ) \,.
\end{align}
\end{subequations}
The backtransformation formally carries a resemblance 
to the Lorentz transformation,
\begin{subequations}
\label{LORVABACK}
\begin{align}
u_0 \, t_A =& \; \gamma_A ( u'_0 \, t'_A - \beta_A \, x' ) \,,
\\[2ex]
x =& \; \gamma_A ( x' - \beta_A u'_0 \, t'_A ) \,.
\end{align}
\end{subequations}
The relativistic factors carry a different functional form,
\begin{equation}
\gamma_A = \sqrt{ 1 + \left( \frac{v_A}{c_0} \right)^2}  \,,
\qquad
\qquad
\beta_A = \frac{v_A}{c_0} \, \sqrt{ 1 + \left( \frac{v_A}{c_0} \right)^2} \,.
\end{equation}
One verifies that
\begin{equation}
x'^2 - {u'_0}^2 \, t_A^2 = x^2 - {u_0}^2 \, t_A^2  \,.
\end{equation}
For the absence of time dilation, one considers events 1 and 2,
with coordinates
\begin{subequations}
\begin{align}
x'_1 =& \; 0 \,, \qquad
t'_{A,1} = 0 \,, \qquad
x_1 = 0 \,, \qquad
t_{A,1} = 0 \,,
\\[2ex]
x'_1 =& \; v_A t_A \,, \qquad
t'_{A,2} = t_A \,, \qquad
x_2 = 0 \,, \qquad
t_{A,2} = \tau \,.
\end{align}
\end{subequations}
One obtains
\begin{equation}
t_{A,2} = \tau =
\left( \gamma_A \, \frac{u'_0}{u_0}  - \frac{v_A^2}{c_0 \, u_0} \right) \, t_A
\end{equation}
There is no time dilation if one chooses the 
parameter $u'_0$ to read as 
\begin{equation}
u'_0 = \frac{c_0 \, u_0 + v_A^2}{c_0 \, \gamma_A} \,.
\end{equation}
If $u_0 = c_0$, then $u'_0 = (1 + v_A^2/c_0^2)^{1/2} \, u_0 = 
\gamma_A \, u_0$.

For comparison (we briefly recall textbook material), 
let us consider the Lorentz transformation,
\begin{subequations}
\begin{align}
c_0 \, t'_E =& \; \gamma_E (c_0 \, t_E + \beta_E \, x ) \,,
\\[2ex]
x' =& \; \gamma_E ( x + \beta_E \, c_0 \, t_E ) \,,
\end{align}
\end{subequations}
where the subscript $E$ stands for the ``electromagnetic''
events according to Vachaspati~\cite{Va2004}.
The backtransformation reads as
$c_0 \, t_E = \gamma_E ( c_0 \, t'_E - \beta_E \, x' ) $
and $x = \gamma_E ( x' - \beta_E \, c_0 \, t'_E )$.
The relativistic factors have the familiar functional form,
\begin{equation}
\gamma_E = \left( 1 - \left( \frac{v_E}{c_0} \right)^2 \right)^{-1/2}  \,,
\qquad
\beta_E = \frac{v}{c_0}  \,,
\end{equation}
One verifies that
$\dd x'^2 - c_0^2 \, {\dd t'_E}^2 = x^2 - c_0^2 \, \dd t_E^2 $.
For the derivation of time dilation, we consider events 1 and 2,
\begin{subequations}
\begin{align}
x'_1 =& \; 0 \,, \qquad
t'_{E,1} = 0 \,, \qquad
x_1 = 0 \,, \qquad
t_{E,1} = 0 \,,
\\[2ex]
x'_1 =& \; v t \,, \qquad
t'_{E,2} = t \,, \qquad
x_2 = 0 \,, \qquad
t_{E,2} = \tau \,,
\end{align}
\end{subequations}
One immediately obtains the familiar time dilation formula,
$t_{E,2} = \tau = t/\gamma_E$.
 
Vachaspati's formalism identifies the ``absolute time'' $t_A$ 
as a formally different parameter from the ``electromagnetic time'' $t = t_E$. 
Furthermore, the ``speed-of-light'' parameter $u_0$ has
to be adjusted for the relative speed  of the primed and unprimed 
coordinate systems.
The relationship of the $u_0$ and $u'_0$ to the 
observed, physical speed of light in both coordinate
systems and the (claimed) 
reconciliation of the Vachaspati transformation
with the Michelson--Morley experiment are discussed in Ref.~\cite{Va2004}.
The Vachaspati transformation reproduces the Galilei 
transformation in the limit $v_A \to 0$ and 
constitutes an alternative to the Lorentz transformation,
with a variable ``speed-of-light'' parameter.
In the primed system, the parameter $u'_0$ can be larger than $c_0$.

The muon lifetime measurement~\cite{RoHa1941},
which demonstrates that fast-moving muons live longer,
is ``reconciled'' in Ref.~\cite{Va2004} with the concept of ``absolute time''
by pointing out that the lifetime of muons is 
determined by the ``electromagnetic time'',
or ``electroweak time'' $t_E$, 
which need to be equal to the ``absolute time'' $t_A$.
It is doubtful if the concept of an ``absolute time''
has any physical interpretation beyond its occurrence
in the Lorentz-like transformation law~\eqref{LORVA}.
However, the Vachaspati transformation 
is indicated here in order to demonstrate that 
a Lorentz-like transformations with 
a variable speed of light parameter $u_0$,
whose value depends on the inertial frame,
have been discussed in the literature.
The Vachaspati transformation is different 
from the modification of the speed of light proposed 
by Franson~\cite{Fr2014} in that the variation of the speed is introduced 
at the classical as opposed to the quantum level.

\end{document}